\documentclass[12pt]{article}
\usepackage{epsfig}
\usepackage{amssymb,amsmath}
\usepackage{epigraph}

\def\la{\langle }
\def\ra{ \rangle }
\def\la{\langle }

\newcommand{\beq}{\begin{equation}}
\newcommand{\eeq}{\end{equation}}
\newcommand{\bea}{\begin{eqnarray}}
\newcommand{\eea}{\end{eqnarray}}

\begin{document}
\renewcommand{\thefootnote}{\fnsymbol{footnote}}

\begin{center}

{\LARGE
Inverse Reinforcement Learning for Marketing   
} 
\vskip1.0cm
{\Large Igor Halperin} \\
\vskip0.5cm
NYU Tandon School of Engineering \\
\vskip0.5cm
{\small e-mail: $igor.halperin@nyu.edu $}
\vskip0.5cm
\today \\
\vskip1.0cm
{\Large Abstract:\\}
\end{center}
\parbox[t]{\textwidth}{
Learning customer preferences from an observed behaviour is an important topic in the marketing literature. Structural models typically model forward-looking customers or firms as utility-maximizing agents whose utility is estimated using methods of Stochastic Optimal Control. We suggest an alternative approach to study dynamic consumer demand, based on Inverse Reinforcement Learning (IRL). We develop a version of the Maximum Entropy IRL that leads to a highly tractable model formulation that amounts to low-dimensional convex optimization in the search for optimal model parameters. Using simulations of consumer demand, we show that observational noise for identical customers can be easily confused with an apparent consumer heterogeneity.
}
 
 \newcounter{helpfootnote}
\renewcommand{\thefootnote}{\fnsymbol{footnote}}
\setcounter{footnote}{0}
 \renewcommand{\thefootnote}{\arabic{footnote}}
\setcounter{footnote}{\thehelpfootnote}

\section{Introduction}

Understanding customer choices, demand and preferences, with customers being consumers or firms, is an eternal theme in the marketing literature. In particular, structural models for marketing build models of consumers or firms by modelling them as utility-maximizing rational agents (see e.g. \cite{Rossi} for a review). Unlike "reduced-form" (purely statistical) models, structural models aim to dissect true consumer choices and demand preferences from effects induced by particular marketing campaigns, thus enabling promoting new products and offers, whose attractiveness to consumer could then be assessed based on the learned consumer utility. 

In particular, in the area of consumer demand research, one can distinguish between static demand and dynamic demand. This paper deals with learning a consumer demand utility function in a dynamic, multi-period setting, where customers  can be both strategic and non-strategic in choosing their optimal consumption over some pre-defined period of time (a week, a month,  a year etc.). Such setting is relevant for marketing different recurrent utility-like plans and services such as cloud computing plans, internet data plans, utility plans (electricity, gas, phone), and so on.  

Structural models approach such problems by modeling forward-looking consumers as rational agents maximizing their streams of expected utilities of consumption over a planning horizon rather than their one-step utility. Structural models typically specify a model for a consumer utility, and then estimate such model using methods of Dynamic Programming and Stochastic Optimal Control. The fact that such models are typically computationally heavy, as they often involve a repeated solution of a Bellman optimality equation, is one of the main stumbling blocks for a wide industrial-level deployment of structural models.

We propose an alternative approach to the problem of learning consumer demand utility in a dynamic, multi-period setting, which is based on Inverse Reinforcement Learning (IRL).  While IRL is widely used in robotics over years \cite{Kober},  more recently it has been applied in other areas as well, in particular to study human behaviour, see e.g. \cite{Krishnan}. We are not aware, however, of any literature that would apply Inverse Reinforcement Learning specifically for problems in marketing. 

The main contribution of this paper is a new version of the Maximum Entropy IRL method (Ziebart 2008), that leads to a highly tractable convex optimization problem for optimal model parameters. Our model enables an easy simulation, which makes it possible to use it to study finite-sample properties of estimators for optimal parameters of the consumer utility. In particular, we use simulations to demonstrate that due to finite-sample effects, consumers with {\it identical} demand utilities might easily be mistaken for {\it heterogeneous} consumers.     
   
\subsection{Related work}
\label{related_work}

Learning customers preferences given their observed behaviour is an active research topic for psychology, marketing, statistical decision making, optimal control, and Artificial Intelligence (AI) communities. Depending on the field, it is usually referred to as the problem of customer choice in the marketing and psychology literature, preference elicitation in the statistical decision making literature, and Inverse Reinforcement Learning in the AI literature. In the specific context of learning the consumer dynamic demand utility, previous research largely follows the Stochastic Optimal Control (SOC) approach. In particular, a 
recent paper by Xu et. al.  \cite{Xu} develops a structural SOC-based model for estimation of mobile phone users' preferences using their observed data daily consumption. 
On the side of Inverse Reinforcement Learning, our framework is rooted in The Maximum Entropy IRL 
(MaxEnt-IRL) \cite{Ziebart_2008, Ziebart_2012} method. 
Other relevant references to the Maximum Entropy IRL are Refs.~\cite{Boularias, Kalakrishnan, Levine_GCS}.

\subsection{Outline of our approach}
\label{sect:outline}

Similarly to Xu {\it et. al.} \cite{Xu}, a framework presented here focuses on consumption data. While our method can be applied to a number of different business settings as outlined in the introduction (such as cloud plans, data plans, utility plans etc.), 
we follow Ref. ~\cite{Xu} and consider a consumption utility of mobile phone users, to facilitate a direct comparison with their approach.

Our model parametrizes users utility (reward) function in terms of a low number of free parameters (that include, in particular, the user price sensitivity), and then estimates these parameters given users' histories of data consumption. 
Unlike Ref.~\cite{Xu}, we do not follow the Stochastic Optimal Control approach, but instead rely on IRL methods developed for similar tasks in the AI and Machine Learning communities. More specifically, we develop a model based on a highly tractable version of a popular Maximum Entropy (MaxEnt) IRL method 
\cite{Ziebart_2008, Ziebart_2012}.

Our approach has a number of important advantages over the model of Ref.~\cite{Xu}. First, our model estimation is much simpler, and amounts to a convex optimization problem with 5 variables, which can be easily handled using standard off-she-shelf optimization software. This enables a very efficient numerical implementation of our model. In contrast, the model of Xu et. al. relies on Monte Carlo for the model estimation. Second, our model is much easier to generalize, if needed, by adding additional features. Third, tractability of our model allows us to investigate the impact of finite-sample "observational noise" on estimated model parameters. This issue was not addressed in Ref.~\cite{Xu} that suggested a substantial users' heterogeneity based on estimation of their model with relatively short (9 months) history for a small number of users. Last but not least, our approach is general enough to be applied, with proper modifications, to learning of customer preferences in other similar settings as suggested above. 

The rest of the paper is organized as follows. In Sect.~\ref{Sect:Model_formulation} we 
present our model. 
In Sect.~\ref{Sect_Counterfactual_simulations}, we show how the estimated model can be 
used for counterfactual simulations and design of marketing strategies. 
Sect.~\ref{Sect:Numerical_experiments} presents numerical experiments. 


\section{Model formulation}
\label{Sect:Model_formulation}

\subsection{User utility function}

Consider a customer that purchased a single-service plan with the monthly price $ F $, initial quota $ q_0 $, and price $p $ 
to be paid for the unit of consumption upon breaching the monthly quota on 
the plan\footnote{For plans that do not allow breaching the quota  $ q_0 $, the 
present formalism still applies by setting the price $ p $ to infinity.}.  
We specify a single-step utility (reward) function of a customer at time $ t = 0, 1, \ldots, T-1 $ (where $ T $ is a lenght of a payment period, e.g. a month) as follows: 

\beq
\label{utility}
r(a_t, q_t, d_t) = \mu a_t - \frac{1}{2} \beta a_t^2 + \gamma a_t d_t  - \eta  p 
( a_t - q_t)_{+}   + \kappa q_t \mathbb{I}_{a_t = 0} 
\eeq  
Here $ a_t \geq 0 $ is the daily consumption on day $ t $, $ q_t \geq 0 $ is the remaining allowance at the 
start of day
$ t $, and  $ d_t $ is the number of remaining days until the end of the billing cycle, and 
we use a short notation $ x_{+} = \max(x,0) $ for any $ x $. 
The fourth term in Eq.(\ref{utility}) is proportional to the payment $ p 
(a_t - q_t)_{+}  $ made 
by the customer once the monthly quota  $ q_0 $ is exhausted.  
Parameter $ \eta $ gives the price sensitivity of the customer,
while parameters $ \mu, \beta, \gamma $ specify the dependence of the user reward on the 
state-action 
variables $ q_t, \, d_t, a_t $. Finally, the last term $  \sim \kappa q_t \mathbb{I}_{a_t = 0}  $ gives the 
reward
received upon zero consumption $ a_t = 0 $ at time $ t $ (here $ \mathbb{I}_{a_t = 0} $ is 
an indicator function that is equal to one if $ a_t = 0 $, and is zero otherwise). 
Model calibration amounts to estimation of parameters $ \eta, 
\mu, \beta, \gamma, \kappa $ given the history of user's consumption. 

Note that the reward (\ref{utility}) can be equivalently written as follows (here $ K = 5 $):
\beq
\label{basis_func_exp}
r(a_t, q_t, d_t) = {\bf \Theta} {\bf \Phi }(a_t, q_t, d_t) = \sum_{k=0}^{K-1} \theta_k \Phi_k(a_t,q_t,d_t)
\eeq
where  
\bea
\label{theta_terms}
& & \theta_0 = \mu \la a_t \ra,  \; 
\theta_1 =  - \frac{1}{2} \beta \la a_t^2 \ra,  \;
\theta_2 =  \gamma  \la a_t d_t \ra ,  \nonumber \\
&&  \theta_3 =  - \eta p  \la ( a_t - q_t)_{+}  \ra , \;  
\theta_4 =   \kappa  \la q_t \mathbb{I}_{a_t = 0}   \ra \nonumber
\eea  
(here $ \la X \ra $ stands for the empirical mean of $ X $), and the following set of basis functions $ \{ \Phi_k \}_{k=0}^{K-1} $ is used:
\bea
\label{basis_functions}
& & \Phi_0(a_t, q_t, d_t) = a_t/ \la a_t \ra,  \nonumber \\
&& \Phi_1(a_t, q_t, d_t) = a_t^2 / \la a_t^2 \ra,  \nonumber \\
& & \Phi_2(a_t, q_t, d_t) = a_t d_t / \la a_t  d_t \ra,  \\
&&  \Phi_3(a_t, q_t, d_t) =  ( a_t - q_t)_{+} / \la ( a_t - q_t)_{+} \ra \nonumber \\ 
&&  \Phi_4(a_t, q_t, d_t) =   q_t \mathbb{I}_{a_t = 0}  / \la q_t \mathbb{I}_{a_t = 0}  \ra \nonumber
\eea  
Our definition of the user reward given by Eq.(\ref{utility}) is similar to the definition proposed in 
Ref.~\cite{Xu}, but differs from it in four aspects. First, we add a possible 
bi-linear dependence of the reward
on daily consumption {\it and} the number $ d_t $ of the remaining days on the plan. Second, we do {\it not} scale parameter
$ \beta $ to $ \beta  = 1 $ as in Ref.~\cite{Xu} (this is because our framework is {\it not}, 
and should not be, scale-invariant). 
Third, we add a reward for zero consumption, given by 
the term $ \kappa q_t \mathbb{I}_{a_t = 0}  $ in Eq.(\ref{utility})\footnote{We note that our specification of no-consumption behavior  is more flexible than what suggested by 
Xu et. al. \cite{Xu}, who interpret zero-consumption events as observations of 
a Gaussian process censored at zero. Instead, our model unties the link between the zero-consumption 
and non-zero consumption events at the price of introducing an additional free parameter $ \kappa $.}. 
Last, and most importantly, 
we do {\it not} add a random daily 
"user shock"  $ \xi_t $ to the value of $ \mu $, as was done in Ref.~\cite{Xu}. 
The fundamental reason for the presence of such "private user shocks" in the user utility function 
in the approach of Ref.~\cite{Xu} is that in the setting of classical Markov decision process (MDP)
problems where the dynamics are typically stochastic but both policy and rewards 
are {\it deterministic}, a demonstrated sub-optimal (instead of an optimal) behavior can
lead to diverging solutions for model parameters, and/or assign zero probabilities to demonstrated 
trajectories\footnote{ The classical MDP problems deal with a 
 completely observed Markov process, where an optimal 
{\it deterministic} policy always exists. Therefore, classical Stochastic Optimal Control methods typically 
work with deterministic policies.}.
 
It is exactly this problem that is addressed by introducing private (i.e. unobserved to the modeller) shocks $ \xi_t $ in the structural model approach adopted in Ref.~\cite{Xu}: the reward function is augmented by an additional random term $ ~ 
\sigma(a_t, q_t, d_t) \xi_t $ with some parametrized "volatility" function $ \sigma(a_t, q_t, d_t) $
(in Ref.\cite{Xu}, the particular form $ \sigma(a_t, q_t, d_t) = a_t $ was used), 
while the exercised policy $ \pi(a_t,q_t,d_t,\xi_t)  $ that gives the next value of $ a_t $ is a 
deterministic function of $ \xi_t $. All model estimations in this framework are therefore based on 
a Monte Carlo simulation of paths of the private shocks $ \xi_t $, and then using them to generate 
paths of the observables $ (a_t, q_t) $ at each time step. 

Instead of relying on the structural models' paradigm, we follow the ideas 
of the Maximum Entropy approach to Inverse Reinforcement Learning (IRL), which is {\it probabilistic} 
and assigns probabilities to observed paths \cite{Ziebart_2008, Ziebart_2012}. 
Due to its probabilistic specification, this approach does {\it not} require introducing a random shock to the utility function in order to reconcile the model with a possible sub-optimal behavior. 
As will be clear below, our approach has a number of advantages over the method of Ref.~\cite{Xu}. Most importantly, it does not need Monte Carlo
to estimate parameters of the user utility, and instead relies on a straightforward Maximum Likelihood Estimation (MLE) with a convex negative log-likelihood function with 5 variables, which can be
done very efficiently using the standard off-the-shelf convex optimization software. Moreover,   
our model enables, if needed, easily generalizations 
of the reward function by adding more of basis functions, while keeping the rest of the methodology intact.

\subsection{Inverse Optimal Control and Inverse Reinforcement Learning}

The problem of estimating the reward (traditionally referred to as the inter-temporal utility function in economics and econometrics literature) given the observed behavior 
is a problem of {\it inverse} optimal control. In {\it direct} optimal control, the objective is to optimize the strategy (i.e. the consumption policy $ \pi(a_t | q_t, d_t) $) such that the expected total reward (total utility) of the user is maximized, assuming the the dynamics of consumption are known, or estimated 
independently. The (direct) Reinforcement Learning (RL) addresses the same problem, but without knowledge of the dynamics and instead relying on samples from the system. In contrast, in the 
Inverse Optimal Control (IOC) or Inverse Reinforcement Learning (IRL) formulations, the problem is to find the reward given the observed behavior (which can be obtained in an off-line or on-line setting).
While in the IOC setting the dynamics are assumed to be known, in the IRL approaches
the dynamics are {\it unknown}, and only {\it samples} obtained under these dynamics are available.

Note that Ref.~\cite{Xu} uses a two-step structural approach to estimating the consumption model 
which first estimates the empirical policy and then finds structural parameters of the utility function that are consistent with this empirically "observed" optimal policy. A similar approach that tries to 
simultaneously estimate both the reward and policy function compatible with this reward is sometimes 
employed in the IRL literature. On the contrary, our Maximum Entropy IRL model is much 
simpler, as in our setting the reward parameters $ \Theta $ automatically fix the user 
policy function, due to the fact that the cumulative consumption process is deterministic given the daily
consumption.

\subsection{Maximum Entropy IRL and Relative Entropy IRL}

The Maximum Entropy IRL (MaxEnt-IRL) \cite{Ziebart_2008, Ziebart_2012}  method is currently the most popular
approach to IRL. The Maximum Entropy argument is applied in our setting to a single-step
(daily) transition probability  $ P \left(q_{t+1},a_t | q_t, d_t \right) $. The MaxEnt solution is
to require that this distribution should match empirical counts $ \phi_k(a_t, q_t, d_t) $ along 
such one-step paths,  but otherwise 
should stay as close as possible to a uniform distribution. Quantitatively, the last condition 
is imposed as the condition of minimization of the Kallback-Leibler relative entropy between the
distribution sought and a uniform distribution.
We use an extension of the MaxEnt-IRL 
called Relative Entropy IRL \cite{Boularias} which replaces the uniform distribution in the MaxEnt
method by a non-uniform benchmark (or "prior") distribution $ \pi_0(a_t | q_t, d_t) $. 
This produces the exponential single-step transition probability:
\bea
\label{single_step_P}
&& P \left(q_{t+1} = q_t - a_t, a_t | q_t,  d_t \right)  \equiv  \pi(a_t | q_t, d_t)  \\
&& = \frac{\pi_0(a_t | q_t, d_t)}{Z_{\theta}(q_t, d_t)} \exp \left( r(a_t, q_t, d_t) \right) = 
 \frac{\pi_0(a_t | q_t, d_t)}{Z_{\theta}(q_t, d_t)} \exp \left(  {\bf \Theta} {\bf \Phi}(a_t, q_t, d_t) \right)
\nonumber
\eea
where  $ Z_{\theta}(q_t, d_t) $ is a state-dependent normalization factor 
\beq
\label{Z_theta}
Z_{\theta}(q_t, d_t) = \int  \pi_0(a_t | q_t, d_t) \exp \left(  {\bf \Theta} {\bf \Phi}(a_t, q_t, d_t) \right) d a_t  
\eeq
We note that most applications of 
MaxEnt IRL deal with multi-step trajectories as prime objects, and define the partition function 
$ Z_{\theta} $ on the space of 
{\it trajectories}. 
While the fist applications of  MaxEnt IRL calculated $ Z_{\theta} $ {\it exactly} for 
small discrete state-action spaces as in \cite{Ziebart_2008}, for large or continuous state-action 
spaces such calculation can only be done {\it approximately} using 
approximate dynamic programming, or other methods. For example, the Relative Entropy IRL approach
of Bourarias et. al. \cite{Boularias}
uses importance sampling from a reference ("background") policy distribution to calculate $ Z_{\theta} $\footnote{Furthermore, the reference distribution can be adapted to the estimated path distribution, as is done  in the Guided Cost Search algorithm of Ref.~\cite{Levine_GCS}.}. It is this calculation that poses the main computational bottleneck for applications of MaxEnt/RelEnt IRL methods  for large or continuous state-action spaces.

In contrast to that, in our approach state-dependent normalization factors 
$ Z_{\theta}(q_t, d_t) $ are defined per each time step. Because we trade a path-dependent
"global" partition function $ Z_{\theta} $ for a local state-dependent factor $ Z_{\theta}(q_t, d_t) $,
we do not need to rely on exact or approximate dynamic programming to calculate this 
factor. Our method is somewhat similar to the approach 
of  Bourarias et. al. (as it also relies on the Relative Entropy minimization), but in our case both the reference distribution
$ \pi_0(a_t | q_t, d_t) $ and normalization factor $ Z_{\theta}(q_t, d_t) $ are defined on a single 
time step, and calculation of $ Z_{\theta}(q_t, d_t) $ amounts to computing the integral 
(\ref{Z_theta}). As we show below, this integral can be calculated analytically with 
a properly chosen distribution $ \pi_0(a_t| q_t, d_t) $. 

To this end, we propose to use a mixture discrete-continuous distribution for the reference ("prior") action distribution $ \pi_0(a_t| q_t, d_t) $:
\beq
\label{prior_mixture_distr}
\pi_0(a_t | q_t, d_t) = \bar{\nu}_0 \delta(a_t) + (1- \bar{\nu}_0)  \tilde{ \pi}_0(a_t | q_t, d_t)  {\mathbb I}_{a_t > 0}
\eeq
where $ \delta(x) $ stands for the Dirac delta-function, and  $  {\mathbb I}_{x> 0} = 1 $ if $ x > 0 $ and zero otherwise. The continuous component $  \tilde{ \pi}_0(a_t | q_t, d_t) $ is given by a spliced Gaussian distribution
\bea
\label{spliced}
\tilde{\pi}_0(a_t | q_t, d_t)  \, = \, \left\{ \begin{array}{clcr}
(1- \omega_0(q_t, d_t)) \phi_1 \left(a_t, \frac{\mu_0 + \gamma_0 d_t}{\beta_0}, \frac{1}{\beta_0} \right) &  \mbox{if $ 0 < a_t \leq q_t $}  \\ 
\omega_0(q_t, d_t) \phi_2 \left( a_t, \frac{\mu_0 + \gamma_0 d_t - \eta_0 p}{\beta_0}, \frac{1}{\beta_0} \right)  &  \mbox{if $ a_t \geq q_t$} \\  
\end{array} \right. 
\eea  
where $\phi_1(a_t,\mu_1,\sigma_1^2) $ and $ \phi_2(a_t, \mu_2, \sigma_2^2) $ 
are probability density functions of two
truncated normal distributions defined separately 
for small and large daily consumption levels, $ 0 \leq a_t \leq q_t $ 
and $ a_t \geq q_t $, respectively (in particular, they
both are  separately
normalized to one). The mixing 
parameter $ 0 \leq \omega_0(q_t, d_t) \leq 1 $ is determined by the continuity condition at 
$ a_t = q_t $:
\beq
\label{junction}
(1 - \omega_0(q_t, d_t)) \phi_1 \left(q_t,  \frac{\mu_0 + \gamma_0 d_t}{\beta_0}, 
\frac{1}{\beta_0} \right) = \omega_0(q_t, d_t) \phi_2 \left(q_t, \frac{\mu_0 + \gamma_0 d_t - \eta_0 p}{\beta_0}, 
\frac{1}{\beta_0} \right) 
\eeq
As this matching condition may involve large values of $ q_t $ where the normal distribution would 
be exponentially small, in practice it is better to use it by taking logarithms of both sides:
\beq
\label{junction_2}
\omega_0(q_t, d_t) = \frac{1}{1 + \exp\left\{ \log \phi_2 \left(q_t, \frac{\mu_0 + \gamma_0 d_t - \eta_0 p}{\beta_0}, 
\frac{1}{\beta_0} \right) -
\log \phi_1 \left(q_t,  \frac{\mu_0 + \gamma_0 d_t}{\beta_0}, 
\frac{1}{\beta_0} \right) \right\} } 
\eeq
The "prior" mixing-spliced distribution (\ref{prior_mixture_distr}), albeit represented in 
terms of simple distributions, leads to potentially quite complex dynamics that make intuitive sense and 
appear largely consistent with observed patterns of consumption. 
In particular, note that Eq.(\ref{spliced}) indicates that large fluctuations $ a_t > q_t $ are centered 
around a smaller mean value 
$ \frac{\mu- \gamma d_t - \eta p}{\beta} $ than the mean value 
$ \frac{\mu- \gamma d_t}{\beta} $ of smaller fluctuations $0 < a_t \leq q_t $. 
Both a reduction of the mean upon breaching the 
remaining allowance barrier and a decrease of the mean of each component with time appear quite intuitive in the current context. As will be shown below, a "posterior" distribution $ \pi(a_t | q_t, 
d_t) $ inherits these properties, while also further enriching the potential complexity of  
dynamics\footnote{in particular, it promotes a static mixing coefficient $ \nu_0 $ to a state- and time-dependent 
variable $ \nu_t = \nu(q_t, d_t) $.}.

The advantage of using the mixed-spliced reference distribution (\ref{prior_mixture_distr}) as 
a reference distribution $ \pi_0(a_t | q_t, d_t) $ is that the state-dependent normalization constant
$ Z_{\theta}(q_t, d_t) $ can be evaluated exactly with this choice:
\beq
\label{Z}
Z_{\theta}(q_t, d_t)  = \bar{\nu}_0 e^{ \kappa q_t} + (1 - \bar{\nu}_0) \left(I_1(\theta, q_t, d_t) 
+ I_2(\theta,q_t, d_t) \right)
\eeq
where
\bea
\label{I_12}
I_1(\theta,q_t, d_t) &=&
(1-\omega_0(q_t, d_t)) \sqrt{\frac{\beta_0}{\beta_0 + \beta}}
\exp\left\{ \frac{(\mu_0 + \mu + (\gamma_0 + \gamma) d_t)^2}{2(\beta_0 + \beta)} 
- \frac{(\mu_0 + \gamma_0 d_t)^2}{2 \beta_0} \right\}  \nonumber \\
 && \times \frac{       
 N \left( - \frac{ \mu_0 + \mu + (\gamma_0 + \gamma) d_t - (\beta_0 + \beta) q_t}{\sqrt{\beta_0 + 
\beta}} \right) - 
N \left( - \frac{ \mu_0 + \mu + (\gamma_0 + \gamma) d_t }{\sqrt{\beta_0 + 
\beta}} \right) 
}{
N \left( - \frac{\mu_0 + \gamma_0 d_t - \beta_0 q_t}{\sqrt{\beta_0}} \right)
- N \left( - \frac{\mu_0 + \gamma_0 d_t }{\sqrt{\beta_0}} \right) 
}
\nonumber  \\
I_2(\theta,q_t, d_t)  &=& 
\omega_0(q_t, d_t) \sqrt{\frac{\beta_0}{\beta_0 + \beta}} 
\exp\left\{ \frac{(\mu_0 + \mu - (\eta_0 + \eta) p + (\gamma_0 + \gamma) d_t)^2}{2(\beta_0 + \beta)} 
\right. \\
&-& \left. \frac{(\mu_0 - \eta_0 p + \gamma_0 d_t)^2}{2 \beta_0} + \eta p q_t \right\}  
  \times \frac{       
 1 - 
N \left( - \frac{ \mu_0 + \mu - (\eta_0 + \eta) p + (\gamma_0 + \gamma) d_t - (\beta_0 + \beta) q_t}{\sqrt{\beta_0 + 
\beta}} \right) 
}{
1
- N \left( - \frac{\mu_0 - \eta_0 p + \gamma_0 d_t - \beta_0 q_t }{\sqrt{\beta_0}} \right) 
} \nonumber 
\eea
where  $ N(x) $ is the cumulative normal probability distribution.

Probabilities of $ T $-steps paths 
$ \tau_i = \left\{ a_t^i, q_t^i, d_t^i \right\}_{t=0}^{T} $   (where $ i $ enumerates different user-paths)
are obtained as products of single-step probabilities:
\beq
\label{multi_step_P}
P(\tau_i) = \prod_{ (a_t, q_t, d_t) \in \tau_i}   
\frac{\pi_0(a_t | q_t, d_t)}{Z_{\theta}(q_t, d_t)} \exp \left(  {\bf \Theta} {\bf \Phi}(a_t, q_t, d_t) \right)
\sim exp \left( {\bf \Theta}  {\bf \Phi}^{(\tau_i)}(a_t, q_t, d_t)
\right)
\eeq
Here $ {\bf \Phi}^{(\tau_i)} (a_t, q_t, d_t)  = \left\{ \Phi_k^{(\tau_i)}  (a_t, q_t, d_t) \right\}_{k=0}^{K-1} $
are cumulative feature counts along the observed path $ \tau_i $:
\beq
\label{cumul_counts}
\Phi_k^{(\tau_i)}  (a_t, q_t, d_t) = \sum_{(a_t, q_t, d_t) \in \tau_i}  \Phi_k (a_t, q_t, d_t)
\eeq
Therefore, the total path probability in our model is exponential in the total reward along a 
trajectory, as in the "classical" MaxEnt IRL approach \cite{Ziebart_2008}, while the pre-exponential
factor is computed differently as we operate with one-step, rather than path-probabilities.  
 
Parameters $ {\bf \Theta} $ defining the exponential path probability distribution (\ref{multi_step_P}) can be estimated by the standard Maximum Likelihood Estimation (MLE) method.
Assume we have $ N $ historically observed single-cycle consumption paths, 
and assume these path probabilities are independent\footnote{A more complex case of co-dependencies between rewards for individual customers can be considered, but we will not pursue 
this approach here. Note that this specification formally enables calibration at the level of an individual customer, 
in which case $ N $ would be equal to the number of consumption cycles 
observed for this user. However, in practice the feasibility of single-name calibration depends on
finite-sample properties of the MLE, which will be discussed in Sect.~\ref{Sect:Numerical_experiments}.}. The total likelihood of observing these data is
\beq
\label{LL}
L (\theta) = \prod_{i=1}^{N} \prod_{ (a_t, q_t, d_t) \in \tau_i}   
\frac{\pi_0(a_t | q_t, d_t)}{Z_{\theta}(q_t, d_t)}
\exp \left(  {\bf \Theta} {\bf \Phi}(a_t, q_t, d_t) \right)
\eeq
The negative log-likelihood is therefore, after omitting the term 
$ \log \pi_0(a_t | q_t, d_t) $ that does not depend on $ \Theta $\footnote{Note that 
$ Z_{\theta}(q_t, d_t) $ still depends 
on $ \pi_0(a_t | q_t, d_t) $, see Eq.(\ref{Z_theta}).}, and rescaling by $ 1/N $,
\bea
\label{neg_LL}
- \frac{1}{N} \log L(\theta) &=& \frac{1}{N} \sum_{i=1}^{N}  \left( 
\sum_{(q_t, d_t) \in \tau_i} \log Z_{\theta}(q_t, d_t)
- \sum_{(a_t, q_t, d_t) \in \tau_i} 
 {\bf \Theta} {\bf \Phi} (a_t, q_t, d_t) \right) \nonumber \\  
 &=& \frac{1}{N} \sum_{i=1}^{N}  \left( 
\sum_{(q_t, d_t) \in \tau_i} \log Z_{\theta}(q_t, d_t) 
- {\bf \Theta} {\bf \Phi}^{(\tau_i)} (a_t, q_t, d_t) \right) 
\eea
Given an initial guess for the optimal parameter $ \theta_k^{(0)} $, we can also consider 
a regularized version of the negative log-likelihood:
\beq
\label{neg_LL_reg}
- \frac{1}{N} \log L(\theta) = \frac{1}{N} \sum_{i=1}^{N} \left( 
\sum_{(q_t, d_t) \in \tau_i} \log Z_{\theta}(q_t, d_t) 
- {\bf \Theta} {\bf \Phi}^{(\tau_i)} (a_t, q_t, d_t)  \right)
+ \lambda || \theta - \theta^{(0)} ||_{q}
\eeq
where $ \lambda $ is a regularization parameter, and $ q = 1 $ or $ q = 2 $ stand for the $ L_1 $- and 
$ L_2 $-norms, respectively. The regularization term can also be given a Bayesian interpretation
as the contribution of a prior distribution on $ \theta_k $ when the MLE estimation (\ref{LL}) is replaced 
by a Bayesian Maximum A-Posteriori (MAP) estimation.  

As is well known, exponential models like (\ref{multi_step_P}) give rise to 
convex negative log-likelihood functions, therefore our final objective function (\ref{neg_LL_reg}) 
is {\it convex} in parameters $ \Theta $ (as can also be verified by a direct calculation), and therefore has a unique solution for any value of 
$ \theta^{(0)} $ and $ \lambda $. This ensures stability of the calibration procedure and a smooth evolution of estimated model parameters $ \Theta $ between individual customers or between groups of 
customers. 

\subsection{Computational aspects}

The regularized negative log-likelihood function (\ref{neg_LL_reg}) can be minimized using 
a number of algorithms for convex optimization. If $ \lambda = 0 $ (i.e. no regularization is used), 
or $ q = 2 $, the objective function is differentiable, and gradient-based methods can be used 
to calibrate parameters $ \theta_k $. 
When $ \lambda > 0 $ and the $ L_1 $-regularization is used, the objective function is non-differentiable
at zero, which can be addressed by using the Orhant-Wise variant of the L-BFGS algorithm, as 
suggested in Ref.~\cite{Kalakrishnan}.

\subsection{Possible extensions for different payment schemes}

So far, we assumed a pricing scheme where a customer pays an upfront price $ F_j $ 
in the beginning of a month, has an initial quota of $ q_{0j} $,, and 
pays a fixed price $ p $ for a unit of consumption once the quota is spent before the end of the month. In practice, there may 
exist a number of modifications to such pricing scheme. First, some service providers may not allow for additional consumption beyond the quota $ q_0 $, so that customers who breach it would only be given a minimally required level of service, for example a low speed access. As was mentioned above, this case can be treated in our framework by taking the limit $ p \rightarrow \infty $ 
in the formulae above. 

Other service/pricing schemes would require further adjustments to the model. 
In particular, in addition to a "main" monthly plan, customers might have access to different plan adjustments and extensions available once the monthly quota is exhausted. This could be handled in our framework by making the state dynamics fully stochastic, rather than locally deterministic as before: 
\beq
\label{add_ons}
q_{t+1} = (q_t + q_e - a_t)_{+}  \nonumber 
\eeq 
where  $ q_e $  is an extra quota that can be added to the plan at a cost $ C(q_{e}) $. As 
such irregular adjustments would probably be only made only a few (or zero) days in a month, 
$ q_e $  would be equal zero at most days in a month, and therefore 
can again be modeled as a mixture of a delta-function at zero and a discrete (or continuous, depending on range of options offered by the service provider) distribution. A mixing weight of this distribution can depend on the current remaining 
quota $ q_t $, remaining days $ d_t $  to the end of a payment period, and possibly some other additional factors. Simultaneously, the reward function  (\ref{utility}) would need to be adjusted by subtracting an additional term  $\eta C(q_{e}) $: 
\beq
\label{utility_2}
r(a_t, q_t, d_t, q_e) = \mu a_t - \frac{1}{2} \beta a_t^2 + \gamma a_t d_t  - \eta  p 
( a_t - q_t - q_e)_{+}   + \kappa (q_t + q_e) \mathbb{I}_{a_t = 0} - \eta C(q_{e})
\nonumber 
\eeq   
While such extensions of the 
model are possible, we leave them for a future research, and concentrate on the basic setting 
presented above in our numerical experiments below.   

\section{Counterfactual simulations}
\label{Sect_Counterfactual_simulations}

\subsection{Action probabilities}
\label{sect:action_probs}

After the model parameters $ \Theta $ are estimated using the MLE method of Eq.(\ref{neg_LL}) 
or (\ref{neg_LL_reg}), the model can
be used for counterfactual simulations of total user rewards assuming that users adopt  
plans with different upfront premia $ F_j  $, prices $ p_j $, and initial quota $ q_j(0) $.
To this end, note that given the daily consumption $ a_t $ and the previous values $ q_{t-1}, d_{t-1}
$, the next values are deterministic: $ q_t = (q_{t-1} - a_t)_{+}, \, d_t = d_{t-1} - 1 $.
Therefore, in our model path probabilities are solely defined by action probabilities, and 
the probability density of different actions $ a_t \geq 0 $ at time $ t $ can be obtained from a one-step 
probability $ P(\tau) \sim \exp\left( r(a_t, q_t, d_t) \right) $. Using Eqs.(\ref{utility}) and 
(\ref{single_step_P}), this gives:
\beq
\label{pi_a}
\pi(a_t | q_t, d_t) = 
 \frac{\pi_0(a_t | q_t, d_t)}{Z_{\theta}(q_t, d_t)}   
\exp\left\{ \mu a_t - \frac{1}{2} \beta a_t^2 + \gamma a_t d_t  - \eta 
p ( a_t - q_t)_{+}   + \kappa q_t \mathbb{I}_{a_t=0}   \right\} 
\eeq 
Using the explicit form of a mixture discrete-continuous 
prior distribution  $ \pi_0(a_t | q_t, d_t) $ given by Eq.(\ref{prior_mixture_distr}), we can express
the "posterior" distribution $ \pi(a_t | q_t, d_t) $ in the same form: 
\beq
\label{mixture_distr}
\pi(a_t | q_t, d_t) = \nu_t \delta(a_t) + (1- \nu_t)  \tilde{ \pi}(a_t | q_t, d_t)  {\mathbb I}_{a_t > 0}
\eeq
where the mixture weight becomes state- and time-dependent:
\beq
\label{nu_t}
\nu_t =  \frac{ \bar{\nu}_0 \exp\{ \kappa q_t  \}}{Z_{\theta}(q_t, d_t) } 
= \frac{ \bar{\nu}_0 \exp\{ \kappa q_t  \}}{ \bar{\nu}_0 e^{  
\kappa q_t } + (1 - \bar{\nu}_0) \left(I_1(\theta, q_t, d_t) 
+ I_2(\theta,q_t, d_t) \right) }
\eeq
(here we used Eq.(\ref{Z})), and the spliced Gaussian component is
\bea
\label{spliced_posterior}
\tilde{\pi}(a_t | q_t, d_t)  \, = \, \left\{ \begin{array}{clcr}
(1- \omega(\theta, q_t, d_t)) 
\phi_1 \left(a_t, \frac{\mu_0 + \mu + (\gamma_0 + 
\gamma) d_t}{\beta_0 + \beta}, \frac{1}{\beta_0 + \beta} \right) &  \mbox{if $ 0 < a_t \leq q_t $}  \\ 
\omega(\theta, q_t, d_t) 
\phi_2 \left( a_t, \frac{\mu_0 + \mu - (\eta_0 + \eta) p + (\gamma_0 + \gamma) d_t }{
\beta_0 + \beta}, \frac{1}{\beta_0 + \beta} \right)  &  \mbox{if $ a_t \geq q_t$} \\  
\end{array} \right. 
\eea 
where the weight $ \omega(\theta, q_t, d_t) $ can be obtained using Eqs.(\ref{pi_a}) and (\ref{Z}).
After some algebra, this produces the following formula
\beq
\label{omega_theta}
\omega(\theta, q_t, d_t) = \frac{ I_2( \theta, q_t , d_t)}{
 I_1( \theta, q_t , d_t) +  I_2( \theta, q_t , d_t)}  = \frac{1}{1 + \frac{I_1( \theta, q_t , d_t)}{
 I_2( \theta, q_t , d_t)} }
\eeq
where functions $ I_1( \theta, q_t , d_t) $, $ I_2( \theta, q_t , d_t) $ are defined above in 
Eqs.(\ref{I_12}). The ratio $ I_1( \theta, q_t , d_t)/ I_2( \theta, q_t , d_t) $ can be equivalently
represented in the following form:
\beq
\label{I1_I2}
\frac{I_1( \theta, q_t , d_t)}{
 I_2( \theta, q_t , d_t)}  =  e^{ - p (\eta_0 + \eta) \left( q_t - \frac{\mu_0 + \mu + (\gamma_0
 + \gamma)d_t}{\beta_0 + \beta} \right) - 
 \frac{ p^2 (\eta_0 + \eta)^2}{ 2(\beta_0 + \beta)} }
 \frac{ \int_{0}^{q_t} e^{ - \frac{1}{2} (\beta_0 + \beta) \left( a_t - \frac{ \mu_0 + \mu
 + (\gamma_0 + \gamma) d_t}{\beta_0 + \beta} \right)^2 } d a_t }{
 \int_{q_t}^{\infty} e^{ - \frac{1}{2} (\beta_0 + \beta) \left( a_t - \frac{ \mu_0 + \mu
 + (\gamma_0 + \gamma) d_t - (\eta_0 + \eta) p}{\beta_0 + \beta} \right)^2 } d a_t
 }
 \eeq
It can be checked by a direct calculation that Eq.(\ref{omega_theta}) with
 the ratio  $ I_1( \theta, q_t , d_t)/ I_2( \theta, q_t , d_t) $ given by Eq.(\ref{I1_I2}) coincides with the
formula for the weight that would be obtained from a continuity condition at $ a_t = q_t $ if we 
started directly with Eq.(\ref{spliced_posterior}). This would produce, similarly to Eq.(\ref{junction_2}),
\beq
\label{junction_posterior}
\omega_0(q_t, d_t) = \frac{1}{1 + \exp\left\{ \log \phi_2 \left(q_t, \frac{\mu_0 + \mu + 
(\gamma_0 + \gamma) d_t - (\eta_0 + \eta) p}{\beta_0 + \beta}, 
\frac{1}{\beta_0 + \beta} \right) -
\log \phi_1 \left(q_t,  \frac{\mu_0 + \mu + (\gamma_0 + \gamma) d_t}{\beta_0 + \beta}, 
\frac{1}{\beta_0 + \beta} \right) \right\} } 
\eeq
The fact that two expressions (\ref{omega_theta}) and (\ref{junction_posterior}) coincide means
that the "posterior" distribution $ \pi(a_t | q_t, d_t) $ is continuous at $ a_t = q_t $ as long as the prior distribution $ \pi_0(a_t | q_t, d_t) $ is 
continuous there. Along with continuity at $ a_t = q_t $, the optimal (or "posterior")
action distribution $ \pi(a_t | q_t, d_t) $ has the same mixing discrete-spliced Gaussian structure
as the reference ("prior") distribution $ \pi_0(a_t | q_t, d_t) $, while mixing weights, means and variances of the component distributions are changed. Such structure-preserving property of our model is similar in a sense to a structure-preservation property of conjugated priors in Bayesian analysis. 
Note that simulation from the spliced Gaussian distribution (\ref{spliced_posterior}) is only slightly 
more involved than simulation from the standard Gaussian distribution. This involves first 
simulating a component of the spliced distribution, and then simulating a truncated normal random variable from this distribution. 
Different consumption paths
are obtained by a repeated simulation from the mixing distribution (\ref{mixture_distr}), along with deterministic updates 
of the state variables $ q_t, d_t $.  Examples will be presented below in Sect.~\ref{Sect:Numerical_experiments}. 

\subsection{Total expected utility of a plan}

Given a consumption plan for a particular service with the monthly fee $ F_j $, initial allowance $
q_{0,j} $ and price $ p_j $ (where $ j = 1, \ldots, J$), the total expected utility at the start of the plan is 
\beq
\label{total_utility}
R_j^{tot} = - \eta F_j + \sum_{t} \mathbb{E} \left[  r(a_t, q_t, d_t) | q_0 = q_{0,j}, p=p_j \right]
\eeq 
If a given customer picked plan $ j $ among all $ J $ possible plans, and we assume the customer acted rationally, this produces 
a set of inequalities
\beq
\label{inequality}
R_j^{tot} \geq R_k^{tot}, \; \;  \forall k \neq j
\eeq
which equivalently can be represented as a set of inequalities for parameter $ \eta $:
\beq
\label{inequality_2}
\eta \geq \frac{\sum_{t} \mathbb{E} \left[  r(a_t, q_t, d_t) | q_0 = q_{0,k}, p=p_k \right] 
- \sum_{t} \mathbb{E} \left[  r(a_t, q_t, d_t) | q_0 = q_{0,j}, p=p_j \right]}{F_k - F_j}, \; \forall k \neq j
\eeq
Depending 
on the specification of a consumption plan, this relation can be used for either a verification (or improvement) of an estimate of $ \eta $ obtained from the MLE procedure presented above, or
alternatively as the only source for calibration of $ \eta $. 
In particular,  some service providers do not permit any additional consumption beyond the plan limit. While this can formally be represented as a particular case of the formalism presented above in the limit $ p \rightarrow \infty $, 
it also would mean that any consumption beyond a quota limit would not be present in data, and hence the price sensitivity 
parameter $ \eta $ could not be learned from the MLE procedure. The only way to infer $ \eta $ 
in such case would be to rely on inequalities (\ref{inequality_2}) which provide a low bound for this parameter. Expected rewards that appear in the numerator of Eq.(\ref{inequality_2}) should be calculated by simulation after other model parameters are estimated using the MLE. 

\subsection{Counterfactual simulation for promotion design}

For the promotion design of possible upgrade offers
$ \left( F_j, q_{0,j}, p_j \right) $ ($ j = 1, \ldots, J $), 
one can rely on a counter-factual analysis of a future customer behavior upon different 
plan upgrade scenarios. We can simulate $ N $ future consumption paths, and then 
again use Eq.(\ref{total_utility}) which we 
repeat here for convenience, to calculate the expected utility of {\it future} consumption:
\beq
\label{total_utility_2}
R_j^{tot} = - \eta F_j + \sum_{t} \mathbb{E} \left[  r(a_t, q_t, d_t) | q_0 = q_{0,j}, p=p_j \right]
\eeq 
Different consumption plans can therefore be compared quantitatively, and sorting them in the decreasing order in $  R_j^{tot} $ (with $j = 1, \ldots, J $) reveals their decreasing attractiveness 
to the customer in terms of their total expected utilities.
 
\section{Numerical experiments}
\label{Sect:Numerical_experiments} 

\subsection{Simulation of consumptions paths}

To test our model, we generate artificial data from the model by simulating paths of daily consumption from 
the mixing-spliced distribution Eq.(\ref{mixture_distr}) as described at the end of Sect.~\ref{sect:action_probs}. 

\begin{figure}[ht]
\begin{center}
\includegraphics[
width=86.92mm,
height=64.04mm]{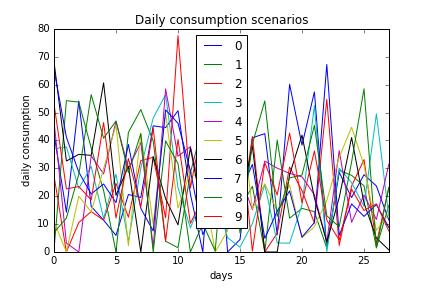}

\caption{Simulated daily consumption} 
\label{fig:daily_consumption_sim}
\end{center}
\end{figure}

\begin{figure}[ht]
\begin{center}
\includegraphics[
width=86.92mm,
height=64.04mm]{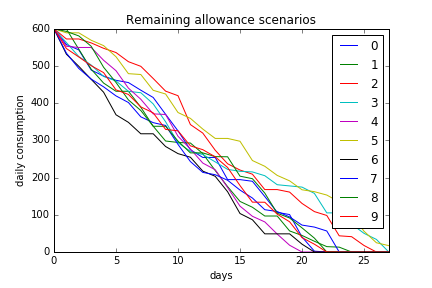}
\caption{Simulated remaining allowance} 
\label{fig:remaining_allowance_sim}
\end{center}
\end{figure}

The simulation of daily consumption is illustrated in Fig.~\ref{fig:daily_consumption_sim}, while the resulting trajectories for remaining allowance are shown in Fig.~\ref{fig:remaining_allowance_sim}, where we 
pick the following values of model parameters:  
$ q_0 =  600$, $ p= 0.55 $, $ \mu = 0.018, \beta  = 0.00125, \gamma = 0.0005,
\eta = 0.1666,  \kappa =  0.0007 $. In addition, we set $ \mu_0 = \mu, \beta_0 = \beta, 
\gamma_0 = \gamma, \eta_0 = \eta, \kappa_0 = \kappa $, and  $ \nu_0 = 0.05 $.

Note that consumption may vary quite substantially from one month to another (e.g. a customer can run out of 
the quota at about  80\% of the time period, or can have a residual unused quota at the end of the month) purely due to the observational noise, even though the utility function stays the same. 


\subsection{Finite-sample properties of MLE estimators}

While the Maximum Likelihood Estimation (MLE) is known to provide asymptotically unbiased 
results for estimated model parameters, in practice we have to deal with data that have limited 
length of history at the level of individual customers. For example, the structural model of Ref.~\cite{Xu} was 
trained on 9 months of data for 1000 customers. While the number of customers to be included for 
analysis can potentially be increased by collecting more data, collecting long {\it individual-level} consumption histories might be more difficult due to a number of factors such as e.g. a customer mobility.

In view of such potential limitations with availability of long time series for service consumption, it 
is important to investigate finite-sample properties of the MLE estimators in the setting of our model.
In particular, note that even if two customers have the {\it same} "true" model parameters, their finite-sample MLE estimates would be in general different for these customers. 

Therefore, ability of the model to differentiate between individual customers hinges upon the size of the bias and variance of its MLE 
estimator in realistically expected settings regarding the amount of data available for analysis.
We note that the authors of Ref.~\cite{Xu} reported a substantial heterogeneity of estimated 
model parameters for their dataset of 9 months of observations for 1000 users, however they did not 
address the finite-sample properties of their estimators, thus leaving out a simplest interpretation of their results as due to "observational noise" in their estimators that would be observed even for a perfectly 
homogeneous set of customers.

We have estimated the "empirical" distribution of MLE estimators for our model by repeatedly
sampling $ N_{m} $ months of consumption history, which is done $ N_p $ times, while keeping 
the model parameters fixed as per above. For each model parameter, we compute a histogram 
of its $ N_p $ estimated values. 

\begin{figure}[ht]
\begin{center}
\includegraphics[
width=160mm,
height=180mm]{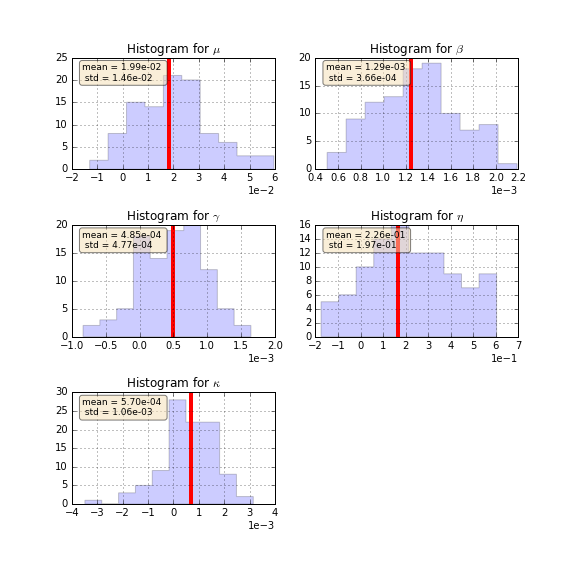}

\caption{Distributions of MLE estimators for $ N_m = 10 $ months of data} 
\label{fig:Hist_10_step_100_sim}
\end{center}
\end{figure}

\begin{figure}[ht]
\begin{center}
\includegraphics[
width=160mm,
height=180mm]{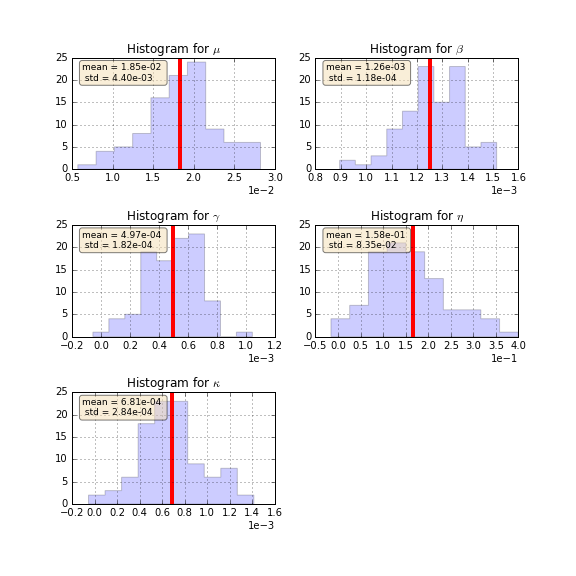}
\caption{Distributions of MLE estimators for $ N_m = 100 $ months of data} 
\label{fig:Hist_100_step_100_sim}
\end{center}
\end{figure}

\begin{figure}[ht]
\begin{center}
\includegraphics[
width=160mm,
height=180mm]{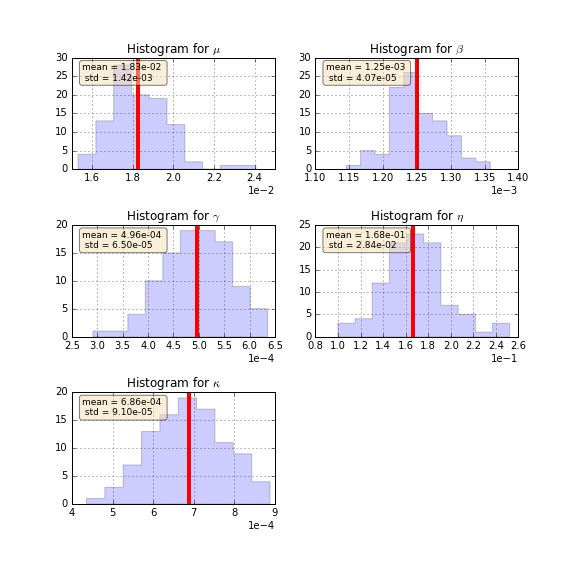}
\caption{Distributions of MLE estimators for $ N_m = 1000 $ months of data} 
\label{fig:Hist_1000_step_100_sim}
\end{center}
\end{figure}

The results are presented in Figs.~\ref{fig:Hist_10_step_100_sim}-
\ref{fig:Hist_1000_step_100_sim}, where we show the resulting histograms for 
$ N_m = 10, 100 $ and 1000 months of data, respectively, while keeping the number of experiments $ N_p = 100 $ for all graphs. Note that for all parameters except $ \beta $, the standard
deviation of the MLE estimate for $ N_m = 10 $ is nearly equal its mean. This implies that 
two users with 10 months of daily observations can hardly be differentiated by the model unless 
their implied parameters differ by a factor of two or more. 

This might cast some doubts on a 
model-implied customer heterogeneity suggested in a similar setting in Ref.~\cite{Xu}, and suggests
that some, if not all, of this heterogeneity can simply be explained by a {\it finite-sample noise} of a 
model estimation procedure, while all customers are actually undistinguishable from the model 
perspective.

On the other hand, one can see how both the bias and variance of 
MLE estimators decrease, as they should, with an increased span of the observation period from 10 user-months to 1000 user-months.  These results suggest that in practice, the model should be 
calibrated using groups of customers with a similar consumption behavior. While the problem of 
finding such groups is outside of scope of this work, this task can be addressed using available 
techniques for clustering time series.  

\section{Summary}

We have presented a very tractable version of Maximum Entropy Inverse Reinforcement Learning (IRL) 
for a dynamic consumer demand estimation, that can be applied technique for designing appropriate marketing strategies for new products and services.  The same approach can be applied, upon proper modifications to similar problems in marketing and pricing of recurrent utility-like services such as cloud plans, internet plans, electricity and gas plans, etc. The model enables easy simulations, which is helpful for conducting counter-factual experiments.
On the IRL/Machine Learning side, unlike most of other versions of the Maximum Entropy IRL, our model does not have to solve
a Bellman optimality equation even once. The model estimation in our approach amounts to convex optimization in a low-dimensional space, which can be solved using a standard off-the-shelf optimization software.  This is much easier computationally than structural models that typically rely on a Monte Carlo simulation for model parameter estimation.


\def\thesection{A}	
\setcounter{equation}{0}
\def\theequation{\thesection.\arabic{equation}}

\end{document}